\documentclass[a4paper,twoside,onecolumn,final,11pt]{article}
\usepackage{natbib}

\usepackage{epsfig}
\usepackage{alltt}
\usepackage{amssymb,amsmath,amsthm,graphicx,psfrag,float}
\usepackage[a4paper,rmargin=2.5cm,lmargin=2.5cm,twoside]{geometry}
\usepackage{lscape}
\usepackage{longtable}
\usepackage[htt]{hyphenat}
\usepackage{tabularx}
\usepackage{array}
\usepackage{fancyhdr}

\newcommand{\zebra}{\texttt{ZEBRA}}
\newcommand{\zebraV}{\texttt{ZEBRA\hspace{1mm}v1.0}}
\newcommand{\lzebra}{\texttt{\large{}ZEBRA}}

\newcommand{\beqn}{\begin{eqnarray}}
\newcommand{\eeqn}{\end{eqnarray}}
\newcommand{\beq}{\begin{equation}}
\newcommand{\eeq}{\end{equation}}
\newcommand{\bal}{\begin{align}}
\newcommand{\eal}{\end{align}}

\def\deg      {{\ifmmode^\circ\else$^\circ$\fi}} 

\newcommand{\apjs}{{\it APJS}}
\newcommand{\aap}{{\it AAP}}

\newcommand{\apjl}{{\it APJL}}

\theoremstyle{remark}

\fancypagestyle{myfancy}
{
\fancyhf{} 
\fancyhead[RE]{\lzebra{} User manual}
\fancyhead[LO]{\slshape \leftmark}
\fancyhead[RO,LE]{\thepage}

}

\fancypagestyle{myplain}
{
\fancyhf{} 
\fancyhead[LO,RE]{\thepage}

}

\newcolumntype{C}[1]{>{\centering\arraybackslash}m{#1}}

\begin{document}

\begin{titlepage}
\vspace{-1cm}
\begin{tabularx}{150mm}{lXr}
\includegraphics[width=50mm]{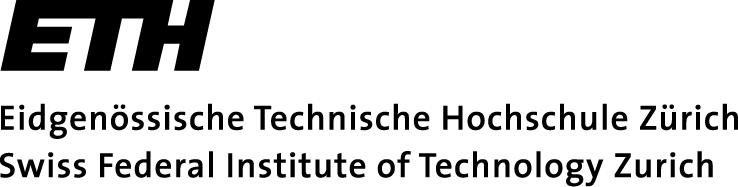} & &
\includegraphics[width=50mm]{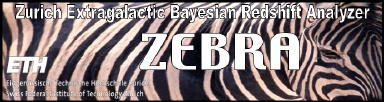} 
\end{tabularx}

\vspace{2cm}

\begin{center}
{\Large The Zurich Extragalactic Bayesian Redshift Analyzer (\lzebra) \\ \vspace{0.4cm} Version 1.0 -- User manual} \\
\vspace{1.5cm}

\texttt{http://www.exp-astro.phys.ethz.ch/ZEBRA} \\
February 1st, 2008 \\
\vspace{1cm}  
\end{center}

\vspace{1cm}

\zebra{}, the \textbf{Z}urich \textbf{E}xtragalactic \textbf{B}ayesian \textbf{R}edshift \textbf{A}nalyzer, is a tool for estimating redshifts and template 
types of galaxies using medium- and broad-band photometric data. \zebra{} employs novel techniques within the 
template-fitting approach to produce high-quality Maximum-Likelihood and Bayesian redshift estimates.

This manuscript serves as a user guide to \zebraV{}. It explains how to use \zebraV{}, specifies
input and output formats, and gives a short account of the available options. The principles
behind \zebraV{} are explained in the reference publication \cite{feldmann06}. 

\zebraV{} is a free and open-source software distributed under the GNU Public License 3. It is kindly requested 
that the use of \zebra{} should be acknowledged with an explicit reference 
to \cite{feldmann06} in the bibliographic list of any resulting publication.

Several upgrades are currently being implemented in \zebra{}. An updated documentation will be provided 
at each new release. Any problems, comments and suggestions on the code and the manual should be 
sent via e-mail to zebra@phys.ethz.ch.

\end{titlepage}


\newpage{}

\clearpage{\pagestyle{empty}\cleardoublepage}

\tableofcontents

\clearpage

\pagestyle{myfancy}
\section{Introduction}

\subsection{About \lzebra}

\zebra{} estimates redshifts and template types of galaxies using a photometric catalog. 
It is based on: \\ 
$(1.)$ An automatic iterative technique to correct the original set of galaxy templates to best represent the SEDs 
of real galaxies at different redshifts;  \\  
$(2.)$ A training set of spectroscopic redshifts for a small fraction of the photometric sample
to improve the robustness of the photometric redshift estimates; and \\ $(3.)$
An iterative technique for  Bayesian redshift estimates, which extracts  the full two-dimensional  redshift and 
template likelihood function for each galaxy.

\paragraph{Input}

The user can choose from, or has otherwise to provide, an initial set of templates and filter 
transmission curves, see Table \ref{tab:providedFilters} and Table \ref{tab:providedTemplates}. 
In addition, \zebra{} needs a catalog containing the magnitudes of the individual galaxies in the 
various filter bands (at least 3 filter bands are needed to run \zebra{} in {\it Maximum-Likelihood mode} or {\it photometry-check mode}).
The different input files are described in detail in section \ref{sect:BasicFiles}.

\paragraph{Output}

\zebra{} offers a variety of output information depending on its run mode.

\begin{description}
\item \zebra{} prints on-screen information during its execution. The level of verbosity of this output can be adjusted 
with the \texttt{-v}/\texttt{--verbosity} option.

\item When run in {\it photometry-check mode} \zebra{} returns a catalog with calibrated 
photometry together with detailed information about the applied changes. 

\item In {\it template-optimization mode} \zebra{} returns the corrected templates as wavelength --
flux density (per unit wavelength) tables. 

\item If \zebra{} is run in {\it Maximum-Likelihood mode}, it will return the best fit redshift and template 
type together with their confidence limits estimated from constant $\chi^2$ boundaries. 
Additionally, \zebra{} returns: (i) the minimum $\chi^2$, (ii) the normalization factor of the best fit 
template, (iii) the rest-frame B-band magnitude and (iv) the luminosity distance for the given cosmological Friedmann-Robertson-Walker model.
If specified by the user, \zebra{} provides further information, e.g. the likelihood functions for all 
galaxies in several output formats or the residuals between best fit template magnitude and measured magnitude 
for each galaxy in each filter band.

\item In the {\it Bayesian mode} \zebra{} calculates the 2D-prior in redshift and template space in an 
iterative fashion. This prior (and, if specified, the interim prior of each iteration step) is returned. The 
final prior is used by \zebra{} to compute a posterior for each galaxy. The posterior can be saved as full 
2D-table or in marginalized form. 
\zebra{}'s output includes the most probable redshift and template type for each galaxy as defined by 
(i) the maximum of the posterior or (ii) after marginalizing over templates types and redshifts, respectively. 
The errors are calculated directly from the posterior. 

\item \zebra{} can also be employed to derive template-based k-corrections using the specified templates and 
filters.

\end{description}
All input and output files of \zebra{} are ASCII-files.

\subsection{The software license}
This program is free software; you can redistribute it and/or
modify it under the terms of the GNU General Public License (GPL)
as published by the Free Software Foundation; either version 3
of the License, or (at your option) any later version.

This program is distributed in the hope that it will be useful,
but WITHOUT ANY WARRANTY; without even the implied warranty of
MERCHANTABILITY or FITNESS FOR A PARTICULAR PURPOSE.  See the
GNU General Public License 3 for more details.

A copy of the GPL 3 is provided at the end of this manuscript.

\subsection{Usage and questions}
It is kindly requested that the use of \zebra{} should be acknowledged with an explicit reference to \cite{feldmann06} in 
the bibliographic list of any resulting publication.

Any problems, comments and suggestions should be sent via e-mail 
to zebra@phys.ethz.ch.

\clearpage

\section{Getting started}

\subsection{About this package}
The package contains the following subdirectories and files:
\begin{itemize}
\item \texttt{src}: The source code of \zebraV{}. How to compile \zebra{} from source is described in \ref{sect:Install}. 
\item \texttt{doc}: This user manual.
\item \texttt{scripts}: Bash-Shell scripts for \zebra{} demonstrating its different run modi. We
suggest to start using \zebra{} with the help of these scripts and modify them accordingly. 
More information about the provided scripts can be found in \ref{sect:UseZebra}.
\item \texttt{examples}: These directories contain the needed input files and generated output files by the provided scripts.
\end{itemize}

\subsection{Installing \lzebra}
\label{sect:Install}
This version provides \zebra{} pre-compiled in several binary formats and as source code. The binaries and the source code package 
can be downloaded from the \zebra{} website at
\begin{center}
\texttt{http://www.exp-astro.phys.ethz.ch/ZEBRA}.
\end{center}

We now explain how to install \zebra{} from source code. First you should read the 
\texttt{README} file in the src directory. Then, before you can compile the \zebra{} package you have to make sure that the BLAS library (\texttt{blas}), the Lapack library (\texttt{lapack}), the GNU
scientific library (\texttt{gsl}) and the Lapack++ library (\texttt{lapackpp}) are installed on your system. 

The first two libraries are installed on most systems, but the GNU scientific library and the Lapack++ library
may be missing on your system. If libraries are not installed on your system you can either install them locally, e.g. in your
home directory\footnote{We suggest you install them into \$HOME$/$local}, or you can ask your system administrator to install them in a directory accessible by all users.

After making sure that the aforementioned libraries are present, you may install \zebra{} as follows:  
\begin{enumerate}
\item Type \texttt{./configure --prefix \it{installation-path}} in the top directory of this package. The directory 
{\it installation-path} is some directory for which you have write permissions, e.g. use the local directory \texttt{\$PWD}. If you have installed 
\texttt{lapackpp} in some directory that is not automatically detected by the configure script
you will have to call \texttt{configure} with additional options specifying the directory. Enter \texttt{configure --help} to obtain a list of all 
available installation options. The \texttt{configure} script sets conservative compiler optimizations according to your
hardware. You can compile \zebra{} with your preferred compiler optimizations by running \texttt{configure} with the option
\texttt{CFLAGS=}\textit{your-compiler-optimization}.

\item Type \texttt{make} and \texttt{make install} to compile the source code and to copy the binary into {\it installation-path}\texttt{/bin}.
\end{enumerate}

\subsection{Checking the installation}

If you want to test the binary of \zebra{} enter the \texttt{bin} directory and type
\texttt{\mbox{zebra -h}}. If everything is fine, \zebra{} will print a complete list of its available options. Be aware that not all available options (especially those which are not described in this manual) are necessarily well tested.

\subsection{How to use \lzebra{}}
\label{sect:UseZebra}

\zebra{} is controlled via command-line options or (more conveniently) with the help of wrapper scripts.
A complete list of all available options and their default values can be obtained by calling \zebra{} with the option \texttt{--help} or \texttt{-h}. 

In this section we first demonstrate how \zebra{} is used in conjunction with command-line options. Later on we describe how to use
\zebra{} conveniently with the provided wrapper scripts.

\begin{figure}
\includegraphics[bb=80 110 760 485, width=160mm]{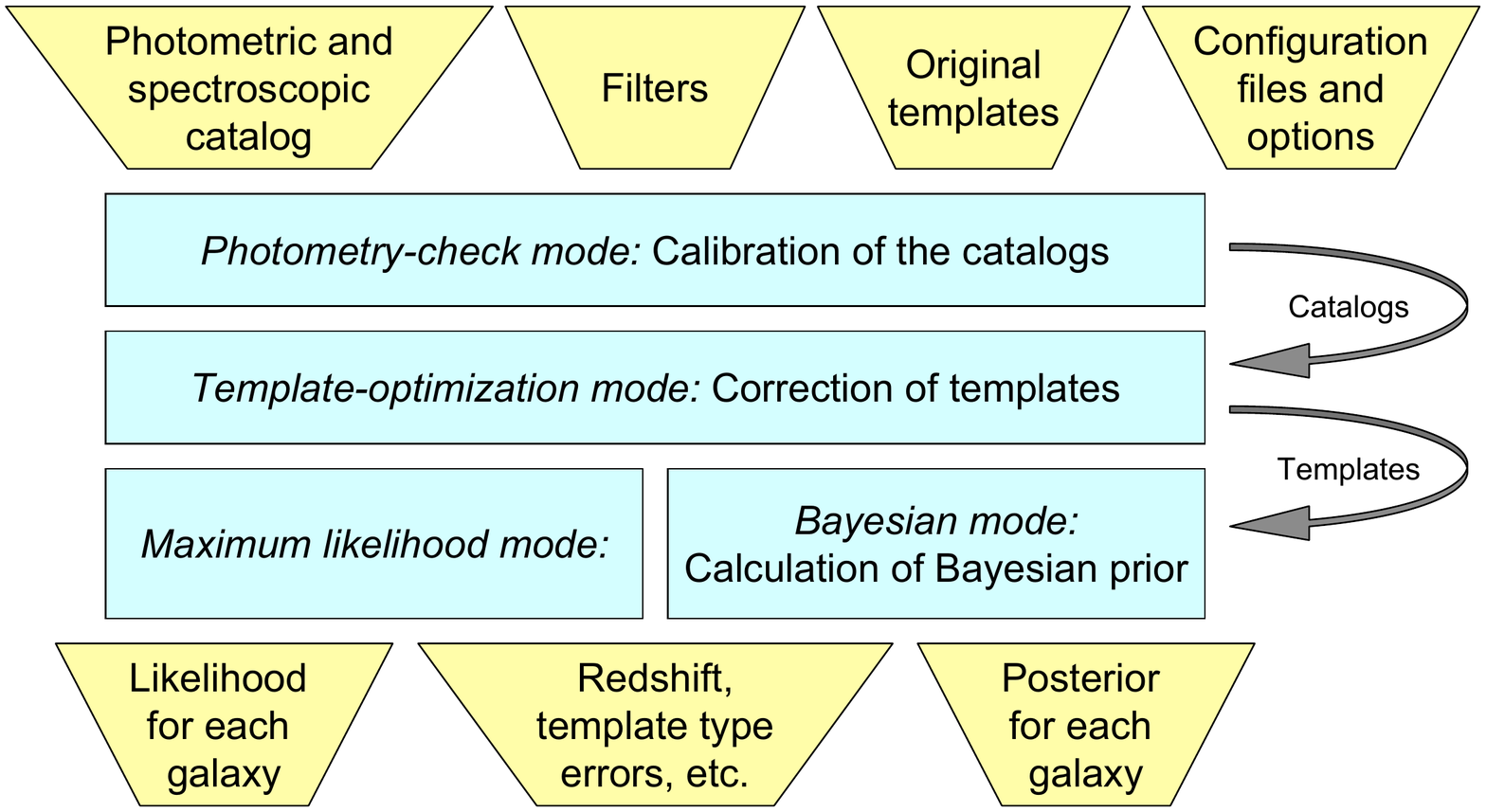}
\caption{The design of \zebra{}. The different operation modi are explained in the text.}
\label{workflow}
\end{figure}

Fig~\ref{workflow} shows the various run modi of \zebra{}. Do not worry of you do not understand all of them at this point.
The run modi are explained in detail in section \ref{sect:runModes}.

\paragraph{Using \zebra{} from the command line:}

Let us have a look at the contents of the directory \texttt{examples/ML\_notImproved}
\begin{alltt}
>ls -1 examples/ML_notImproved/
filter.conf
template.conf
testCatalog.cat
testCatalog.cat.def
\end{alltt}
The files \texttt{filter.conf} and \texttt{template.conf} specify which filters and templates are to
be used by \zebra{}. The filter and template files are stored in \texttt{examples/filters/} 
and \texttt{examples/templates/}. The photometric catalog is denoted \texttt{testCatalog.cat} and has an 
associated definition file \\
\texttt{testCatalog.cat.def}. The latter tells the program which columns 
of the catalog file correspond to which filters, which columns contain
redshift information and also allows one to set minimum errors for photometric fluxes. The content of 
\texttt{testCatalog.cat.def} is shown in fig.~\ref{testCatDef}.

Let us enter the directory \texttt{examples/ML\_notImproved}
\begin{alltt}
>cd examples/ML\_notImproved
\end{alltt}
and call \zebra{} in Maximum-Likelihood mode:
\begin{alltt}
>zebra -c testCatalog.cat -F ../filters -T ../templates -k -i 5 
-z 0 -Z 2 -D 0.005
\end{alltt}
This example assumes that you have included the location of the \zebra{} binary in your \$PATH variable. If not, you should call the executable of \zebra{} by 
its full filename (incl. the path). 

Options are passed to \zebra{} by the common getopt/getoptlong--syntax. This means that you can specify options either with a dash followed by a single letter or a double-dash followed by a keyword.
In our example the first three options denote the catalog file to read, the location 
of the filter transmission curves and the location of the templates. The option \verb|-k| specifies that \zebra{} is run in Maximum-Likelihood mode. The number of log-interpolated templates is set to 5 with the command \verb|-i 5|. The range (from 0 to 2) and
width (0.005) of the redshift grid is set with the options  \verb|-z 0| \verb|-Z 2| and \verb|-D 0.005|, respectively.

Upon successful termination of \zebra{} the file \verb|ML.dat| is created in the current directory. This 
file summarizes all relevant information about the catalog entries, e.g. the best fit redshift and template type. A
detailed header describes all contained columns.

\begin{figure}
\label{testCatDef}
\begin{alltt}
# column_name                 column_number    minimum_magnitude error
#                 type(F,dF,Z)        1-sigma magnitude                    
#
z_cat                   Z       17
z_phot                  Z       18
u_cfht.res              F       1
B.res                   F       3
g.res                   F       5
V.res                   F       7
r.res                   F       9
i.res                   F       11
z.res                   F       13
flamingos_Ks.res        F       15
u_cfht.res              dF      2       x       0.05
B.res                   dF      4       x       0.05
g.res                   dF      6       x       0.05
V.res                   dF      8       x       0.05
r.res                   dF      10      x       0.05
i.res                   dF      12      x       0.05
z.res                   dF      14      x       0.05
flamingos_Ks.res        dF      16      x       0.2
\end{alltt}
\caption{ 
The content (shortened) of the catalog definition file \texttt{testCatalog.cat.def}. The first column associates a name
to each column of the catalog file. If this column contains flux or flux error information the given name is matched 
with the filter names from \texttt{filter.conf}. The second column in \texttt{testCatalog.cat.def} specifies the type of
the column. Here \texttt{F}, \texttt{dF} and \texttt{Z} stand for filter band, filter band error and redshift information, 
respectively. An upper-limit magnitude is specified in the fourth column of the filter band error entries and a minimum magnitude 
error in the fifth column. Missing columns are marked with an \texttt{x}.}
\end{figure}

\paragraph{Using \zebra{} with wrapper-scripts:}

Enter the \texttt{scripts} directory below the top-level directory of the package and open the 
Perl script \texttt{callzebra\_ML\_notImproved}. It sets various path variables, the name of the catalog containing the photometric data, various options and finally calls \zebra{} with all these options. Executing the script by entering \texttt{callzebra\_ML\_notImproved} will 
run \zebra{} and create the output file  \verb|ML.dat| in the directory \texttt{examples/ML\_notImproved} as before.

Each of the provided scripts demonstrates one of the different run modi of \zebra{}. In particular:
\begin{itemize}
\item \texttt{callcatalogCorrect}: Runs \zebra{} in {\it photometry-check mode}.
\item \texttt{callimprove}: Runs the {\it template-optimization mode} of \zebra{}. It creates improved templates.  
\item \texttt{callzebra\_ML\_notImproved}: Runs \zebra{} in {\it Maximum-Likelihood mode} on a catalog consisting of (only) 10 objects using default templates.
\item \texttt{callzebra\_M\_improved}: Same as before but now (previously) improved templates are used. Please note, the ``improved'' templates used in this example
are only toy templates and should not be used for scientific purposes.
\item \texttt{callzebra\_Bayes}: Demonstrates how \zebra{} is run in {\it Bayesian mode}.
\item \texttt{callkCorrect}: Runs \zebra{} in {\it k-Correction mode}.
\end{itemize} 

Please note that all scripts can be executed independently from each other. \textbf{We do not claim that the values provided in the scripts
are necessarily well suited for the provided examples nor for your specific application. We encourage you to test carefully 
which option values are appropriate for your application.}

\section{Basic files and settings}
\label{sect:BasicFiles}

This section describes both the syntax of many options and the layout of the basic files
needed in almost every \zebra{} session. Please note that all input and output files of \zebra{} 
are ASCII files in a column-based, white-space separated table-format. Individual lines may be 
commented out by prepending a hash (\texttt{\#}) sign.

\subsection{Directories}

All directories can be specified with \zebra{}'s basic options (see \ref{sect:BasicOptions}). 
Each unspecified input or output directory defaults to the local directory.

\subsection{The filter files}
\label{sect:filterFiles}
Filter files describe the transmission curve of a filter bands. The first column in each filter file specifies
the wavelength in \AA{} and the second column the photon transmittance (between zero and one).
The wavelength values need to grow strictly monotonically from top to bottom of the
filter file. \zebra{} checks whether this is fulfilled and prints a warning if not.  

This package provides some filter files in the \texttt{examples/filters} directory, see Table \ref{tab:providedFilters}.

\begin{table}
\begin{tabular}{|c|c|}
\hline
file name & filter description \\
\hline\hline
u\_cfht.res & CFHT U*\\
B.res       & Subaru B \\
V.res       & Subaru V \\
g.res       & Subaru g+\\
r.res       & Subaru r+\\
i.res       & Subaru i+\\
z.res       & Subaru z+\\
flamingos\_Ks.res & KPNO/CTIO Ks \\
\hline
\end{tabular}
\caption{The provided filter files in this release (and many more) are publicly available at \texttt{http://www.astro.caltech.edu/$\sim$capak/cosmos/filters}.}
\label{tab:providedFilters}
\end{table}

\zebra{} needs the specification of one designated filter (we will call it B-filter throughout
this manuscript)
\begin{itemize}
\item to ensure an absolute magnitude cut,
\item to output absolute magnitudes,
\item to normalize the templates.
\end{itemize}

The designated filter band \emph{does not need} to be one of the filter bands of the photometric catalog. The only requirement is that the filter file is also present in the filterpath.

The filename of the designated filter can be specified with the \texttt{--b-filter-name} option, the default is \texttt{B.res}.

\subsection{The template files}
\label{sect:templateFiles}

Each template file contains the SED of a real or synthetic galaxy in the following format. The first
column specifies the wavelength in \AA{} and the second the spectral flux density (per unit wavelength 
and with arbitrary scale). The wavelength values need to grow strictly monotonically from top to bottom of the
template file. \zebra{} checks whether this is fulfilled and prints a warning if not. 

This package provides some template files in the \texttt{examples/templates} directory, see Table \ref{tab:providedTemplates}.

Depending on the redshift range of interest you will need to make sure 
that all your templates cover a sufficiently large wavelength range. Otherwise you will lose some
filter bands. The reason is that \zebra{} selects the best fit based on the global minimum 
$\chi^2$ and therefore it can only safely use the filter bands for which fluxes can be calculated 
for all redshifts and all templates. \zebra{} will warn you if your templates need to be extended to shorter wavelength (``UV'') or 
longer wavelength (``IR'').

\begin{table}
\begin{tabular}{|c|c|c|}
\hline
file name & template description & source \\
\hline\hline
El\_cww.sed & elliptical galaxy & \cite{cww} \\
Sbc\_cww.sed       & Sbc galaxy & \cite{cww} \\
Scd\_cww.sed       & Scd galaxy & \cite{cww}\\
Im\_cww.sed       & irregular galaxy & \cite{cww} \\
SB3\_kin.sed       & starbursting galaxy & \cite{kin}\\
SB2\_kin.sed       & starbursting galaxy & \cite{kin}\\
\hline
\end{tabular}
\caption{The provided template files in this release are publicly available as part of the BPZ package (\texttt{http://acs.pha.jhu.edu/$\sim$txitxo/bayesian}).}
\label{tab:providedTemplates}
\end{table}

\subsection{The catalog file}
\label{sect:CatalogFile}
This file contains the photometric data (and spectroscopic redshifts if available) of the galaxies in question. Each row corresponds to an individual galaxy.
The rows contain the photometric magnitudes (must be given in AB magnitudes), corresponding magnitude errors, the redshift (if available) and possible other information, e.g. an ID. The latter information is currently ignored by \zebra{}.

The assignment of the given rows to their filter bands etc. is specified in the catalog definition file, see \ref{sect:StandardConfFiles}.

\zebra{} converts all magnitudes and magnitude errors into flux densities (in erg s$^{-1}$ cm$^{-2}$
Hz$^{-1}$) and \emph{relative} flux density errors according to the following formulae:
\[
  \text{flux} = 10^{  -0.4(\text{mag} +48.6)},  e(\text{flux}) =  e(\text{mag})\log(10.)/2.5
\]

A magnitude entry $\geq{}99$ in a filter band is recognized as too weak a signal to be detected. A magnitude entry $\leq{}-99$ is recognized as not being measured. 

If for a given galaxy no magnitude in a filter band is specified (-99) that filter band is discarded, i.e. not used in the Maximum-Likelihood redshift estimation of that galaxy. If the magnitude is non-detected (99) but a valid magnitude error is given, i.e 0$<$error$<$99, the magnitude is either set to  a provided ``1-$\sigma$-magnitude'' (see \ref{sect:StandardConfFiles}) or otherwise the filter band value is discarded for this galaxy. If the considered filter band contains a valid magnitude ($>-99$ and $<99$) but the error is not valid, the error will be either set to the minimum magnitude error if (i) this error is provided in the catalog configuration file and (ii) the option \texttt{--allow-missing-error} is enabled, or the filter band is discarded otherwise.
If a minimum magnitude error is provided, all magnitude errors in the catalog will be increased if necessary to be at least equal to this minimum magnitude error.

For the benefit of the user \zebra{} displays a warning (and this is all it does) if some magnitude in the catalog is out of a given range. This range (the default is from 5 to 35) can be specified with the \texttt{--mag-catalog-low} and \texttt{--mag-catalog-high} options.

The catalog name has to be supplied to \zebra{} with the \texttt{-c} option.

\subsection{The standard configuration files}
\label{sect:StandardConfFiles}
The configuration files determine (i) which templates and filters are used by \zebra{} and (ii) the structure of the catalog file.

In the following we assume for sake of simplicity that the default file names of the configuration files are used, i.e.
\texttt{filter.conf}, \texttt{template.conf} and \textit{catalog}\texttt{.def}. Here \textit{catalog}
is the name of the photometric catalog.
The names can be changed with the options \verb|-f |\textit{filter-conf-file}, 
\verb|-t |\textit{template-conf-file} and \verb|-d |\textit{catalog-def-file} respectively.

The structure of the configuration files is shown in detail in tab.~\ref{tab:filter.conf}, tab.~\ref{tab:template.conf} and tab.~\ref{tab:catalog.def} and 
will be explained below.

\begin{table}
\begin{tabular}{|c|c|c|c|}
\hline
column & content type & content & remark\\
\hline\hline
1      & string  & filter name & \\
2      & integer & filter used (1) or not used (0) & optional, default: 1 \\
\hline
\end{tabular}
\caption{The structure of the filter configuration file. Each entry has to be written on a single line. Comment
lines have to start with \#.}
\label{tab:filter.conf}
\end{table}

\begin{table}
\begin{tabular}{|c|c|c|c|}
\hline
column & content type & content & remark\\
\hline\hline
1      & string  & template name & \\
2      & integer & template used (1) or not used (0) & optional, default: 1\\
3      & float   & lower redshift limit & optional, default: 0\\
4      & float   & upper redshift limit & optional, default: 10 \\
5      & float   & mass to light ratio in B-band & optional, default: 1\\
6      & string & name of rho-sigma-file & optional, no default\\
\hline
\end{tabular}
\caption{The structure of the template configuration file. Each entry has to be written on a single line. Comment
lines have to start with \#. The lower and upper redshift limits are used to restrict the use of templates to a certain
redshift range (see options \texttt{--chi2-add} and \texttt{--chi2-mult}). However, the use of those options is not recommended. 
The mass-to-light ratio (default 1) is used in the construction of linear interpolated templates. 
The last column may state the name of a file containing $\sigma^2$ and $\rho^2$ values which are used in the {\it template-improvement mode}, see \ref{sect:OtherConfFiles}.}
\label{tab:template.conf}
\end{table}

\begin{table}
\begin{tabular}{|c|c|c|c|}
\hline
column & content type & content & remark\\
\hline\hline
1      & string  & column name & the filter name or \\
       &          &            &  (arbitrary) name for a redshift column\\
2      & string  & column type & one of F, dF, Z, dZ \\
3      & integer & column number & starting with 1\\
4      & float & 1-$\sigma$ magnitude & optional; only for dF; no default\\
5      & float & minimum magnitude error & optional; only for dF; no default\\
\hline
\end{tabular}
\caption{The structure of the catalog configuration file. Each entry has to be written onto a single line. Comment
lines have to start with \#.}
\label{tab:catalog.def}
\end{table}

The first two configuration files specify which filters and templates are available to the program. 

Each row of \texttt{template.conf}, excluding comments, corresponds to a template file. Each row may consist of several columns.
The first column specifies the template name, which is obligatory; the other columns are optional. The second column may contain a 1 (in this case \zebra{} will 
use this template, which is the default) or a 0 (\zebra{} will ignore this template). The 3rd and 4th column are used to specify minimum and maximum ranges of the template. These columns should not be used in this release of \zebra{}  but might be useful in future extensions.
The 5th column can contain a mass-to-light ratio (default 1), which is used in the construction of the linear interpolated templates. The 6th column may contain 
the name of a file with $\rho$ and $\sigma$ values specifically for this template. This file is described in more detail in \ref{sect:OtherConfFiles}. In order to 
address a column without specifying the values of preceding columns the special value ``x'' may be used.

The layout of \texttt{filter.conf} is similar. For each row up to two columns are recognized by \zebra{}. The first column contains the full filter 
name (e.g. \texttt{B.res}), the second column is optional and may contain a 1 (i.e. \zebra{} will use this filter, which is the default) or 0 (i.e. \zebra{} will ignore this filter).
Since the columns of the photometric catalog are defined in the \textit{catalog}\texttt{.def} file it might happen that not all filters in \textit{catalog}\texttt{.def}
are mentioned in \texttt{filter.conf} or not all filters in \texttt{filter.conf} are actually used in \textit{catalog}\texttt{.def}. In these cases \zebra{} will only consider the filters which are mentioned in both files. 

The catalog definition file specifies which columns of the catalog file contain the magnitudes and errors of which filter band.
Each row consists of at least 3 columns. Additional columns carry facultative information. The first column  contains the full 
filter name if this particular row specifies a filter band or a filter band error. Otherwise it might contain an arbitrary string. The third column denotes the
column number in the catalog file to which this row corresponds. The second column
contains one of the following labels: \texttt{F}, \texttt{dF}, \texttt{Z}. According to the provided label \zebra{} decides whether the respective column in the catalog file
corresponds to the magnitude in a filter band, the magnitude error in a filter band or a redshift column. The available options described by the facultative columns depend on the label. In filter bands rows (\texttt{F}) a 4th column is recognized by \zebra{} as specifying a zero-point offset. In filter band errors rows (\texttt{dF}) the 
4th column may contain a ``1-$\sigma$ magnitude'' and the 5th column a minimum magnitude error. Again,  the special value ``x'' may be used to address a column without specifying the values of preceding columns.

The zero-point offset is subtracted from the corresponding filter band magnitude when the catalog is read by \zebra{}.
Please do not rely on this special feature but run \zebra{} in {\it photometry-check mode} (see \ref{sect:PhotometryCheck}) to accurately remove offsets
from the catalog itself. The  1-$\sigma$ magnitude is used as an upper magnitude limit for galaxy entries with ``non-detected'' magnitudes in the catalog file (see \ref{sect:CatalogFile}). The minimum magnitude error 
offers the user the possibility to enforce a lower limit on the \emph{relative} flux errors seen by \zebra{}. This is sometimes necessary because a small flux error in a filter band can strongly influence the outcome of Maximum-Likelihood method. Please note that the photometric errors provided in the catalog usually do not include, e.g., errors due to missing or mismatching templates.  Using sensible errors, however, is very important in order to obtain robust photometric redshifts.

\subsection{Other configuration files}
\label{sect:OtherConfFiles}
We now briefly describe files which are specific to certain run modi of \zebra{}. More information can be found in the 
section discussing the corresponding run mode.

The catalog-correction configuration file (default \texttt{catalogCorrection.def}) is an input file used in the {\it photometry-check mode}.
It gives an account of the filter bands to be corrected and how the correction is done (number of magnitude bins, regression type, limits of correction, \ldots).

The template-improvement configuration file (default \texttt{templateImprovement.def}) is an input file used in the {\it template-improvement mode}.
It specifies the limits of the redshift bins employed in the template improvement.

\subsection{Basic output files}
\label{sect:OutputFiles}

In {\it photometry-check mode}, {\it template-optimization mode} or {\it Maximum-Likelihood mode} \zebra{} produces 
a file with (default) name \texttt{ML.dat}.
In the {\it Bayesian mode} a very similar file with the (default) name \texttt{Bayes\_ML.dat} is created.

\texttt{ML.dat} contains in tabulated form the results of the Maximum-Likelihood fit applied to the catalog entries, while \texttt{Bayes\_ML.dat} contains in a similar fashion the outcome of the Bayesian (maximum posterior) analysis. Likelihood or posterior functions for individual catalog entries can be obtained by enabling the appropriate \texttt{--likelihood} or \texttt{--posterior} option. 

The prefix (\texttt{ML}) of the file \texttt{ML.dat} can be changed with the option \texttt{--outputbase} \textit{filename}. When \zebra{} is run in {\it Maximum-Likelihood mode} the file contains the following information:
\begin{itemize}
\item Each no-comment line corresponds to the same no-comment line from the catalog file, i.e. the results are printed in the order of the input catalog.
\item The spectroscopic redshift and the number of filter bands used in the fit.
\item The set of estimated redshift $z$, template type $t$, normalization (quantity $a$ defined in eq. A1 of \cite{feldmann06}),
 $\chi^2$, absolute magnitude (in the designated filter band, see \ref{sect:filterFiles}), luminosity distance for following cases: (i) the pair $(z,t)$ corresponding to the maximum of the likelihood $L(z,t)$ (i.e. the minimum of $\chi^2(z,t)$), (ii) the pair $(z,t)$ corresponding to the maximum of the marginalized likelihood $L(z)$ and the template which is best fit with that redshift, 
 (iii) the pair $(z,t)$ corresponding to the maximum of the
marginalized likelihood $L(t)$ and the redshift which best fits this template type.
\item Errors for the redshift and template type estimates using (i) constant $\chi^2$ boundaries (ii) percentiles
of the marginalized likelihood functions. 
\end{itemize}

An estimated redshift $z$ of -1 indicates that \zebra{} was not able to perform the $\chi^2$ fit. This might happen if, e.g. not enough filter bands were available or the absolute magnitude cut discarded the catalog entry (option \texttt{--mb-mode 1}). 

If the \texttt{--normalize-templates} option is specified, \zebra{} normalizes all provided
templates to unity flux-density per unit wavelength in the dedicated B-filter. However, 
the use of this option is only necessary if you want to use linearly interpolated templates. The 
conversion factor between original and normalized template is displayed if you run \zebra{} with 
verbosity 2 or higher. You will need to multiply the normalization given in \texttt{ML.dat} with this
conversion factor to obtain the total normalization w.r.t. the original template.

We will now explain how the template number that \zebra{} outputs relates to the input templates provided to \zebra{}.  Let $n$ be the total number of template names (first column) in the template configuration file that are active, i.e. the corresponding second 
column is 1 or is not given (in which case the value defaults 1). We call these templates ``basic templates'' in order to distinguish them from interpolated templates created internally within \zebra{}.
If no interpolated templates are used the template number $t$ which \zebra{} outputs corresponds to the $t+1$-th active template in the template configuration file. Let us now assume that \zebra{} is run with $l\geq{}0$ logarithmically interpolated templates. 
The template types range now from 0 to $(n-1)l+n-1$. This range corresponds to a smooth
sequence of basic and interpolated templates with 0 corresponding to the first basic template and $(n-1)(l+1)$ to the last. 
If type $t$ is a multiple of $l+1$ it corresponds to the $t/(l+1)+1$-th basic template in the template configuration file.
Otherwise $t$ describes a log-interpolated template between the $\lfloor{}t/(l+1)\rfloor{}+1$-th and the $\lceil{}t/(l+1)\rceil{}+1$-th basic template, where $\lfloor{}x\rfloor{}$ denotes the largest integer not exceeding x and $\lceil{}x\rceil{}$ the smallest integer not smaller than x. For example, consider the case that 6 basic templates and 5 log-interpolations between each pair of adjacent basic templates 
are used. Then the template types range from 0 to 30, the template numbers 0, 6, 12, 18, 24 and 30 denote the 6 basic templates and all other template numbers denote interpolated templates.

The infix (\texttt{ML}) of the file \texttt{Bayes\_ML.dat} can be changed with the option \texttt{--outputbase} \textit{filename}. This output file contains the following information.
\begin{itemize}
\item The spectroscopic redshift and the number of filter bands used in the fit.
\item The set of estimated redshift $z$, template type $t$, normalization (quantity $a$ defined in eq. A1 of \cite{feldmann06}),
$\chi^2$, absolute magnitude (in the designated filter band, see \ref{sect:filterFiles}), 
luminosity distance for following cases: (i) the pair $(z,t)$ corresponding to the maximum of 
the posterior $P(z,t)$, (ii) the pair $(z,t)$ corresponding to the maximum of the marginalized 
posterior $P(z)$ and the template which is best fit with that redshift, 
(iii) the pair $(z,t)$ corresponding to the maximum of the
marginalized posterior $P(t)$ and the redshift which best fits this template type.
\item Errors for the redshift and template type estimates using percentiles
of the marginalized (i.e. summing over a parameter) and projected (i.e. fixing a parameter) posteriors.
\end{itemize}

More information about the entries in both output files can be found in the file header.

\subsection{Basic options}
\label{sect:BasicOptions}
The following options are very useful in almost all run modi:
\begin{description}
\item \texttt{--filterconf} \textit{filename}:  Filter-configuration file.
\item \texttt{--filterpath} \textit{pathname}:  Directory containing the defined filters.
\item \texttt{--templateconf} \textit{filename}: Template-configuration file.
\item \texttt{--templatepath} \textit{pathname}: Directory containing defined templates.
\item \texttt{--catalog} \textit{filename}: Filename of the catalog file.
\item \texttt{--catalogpath} \textit{pathname}: Directory containing catalog file.
\item \texttt{--catalogdef} \textit{filename}: Filename of the catalog-configuration file.
\item \texttt{--outputbase} \textit{filename}: Basename (prefix) of the output file.
\item \texttt{--outputpath} \textit{pathname}: Directory to which output is written.
\item \texttt{--zmin} \textit{float}: Lower limit of allowed redshift range.
\item \texttt{--zmax} \textit{float}: Upper limit of allowed redshift range.
\item \texttt{--dz} \textit{float}:  Difference between lowest redshift grid points.
\item \texttt{--log-zBin}: Use redshift grid in log(1+z) instead of binning linear in redshift.
\item \texttt{--ig-absorption} \textit{integer}: No intergalactic absorption (0), absorption according to \cite{1995ApJ...441...18M} (1), 
or according to \cite{2006MNRAS.365..807M} ($>1$), see \ref{sect:IGA}.
\item \texttt{--mb-mode} \textit{integer}: This option specifies how to deal with entries for which the best fitting (redshift, template) pair implies an absolute magnitude
out of range (in the designated filter band, see \ref{sect:filterFiles}). (0) The fit result is used, (1) The fit result is discarded and the catalog entry is marked as non-fittable, (2) The fit result is discarded and the best fitting (redshift,template) pair which obeys the absolute magnitude is used as fit result.
\item \texttt{--mb-low} \textit{float}: Lower (brighter) limit of allowed absolute magnitude.
\item \texttt{--mb-high} \textit{float}: Upper (fainter) limit of allowed absolute magnitude.
\item \texttt{--rightz}: Use spectroscopic redshift (if available).
\item \texttt{--log-interpolation} \textit{integer}: Number of interpolations in magnitude space (logarithmic interpolations) between each adjacent template pair.
\item \texttt{--save-flux}:  The calculated flux densities (per unit frequency) are saved into flux file (-x). The
stored quantity is defined in eq. A3 of \cite{feldmann06}. 
\item \texttt{--load-flux}:  Flux densities (per unit frequency) are read from flux file (-x). The
stored quantity is defined in eq. A3 of \cite{feldmann06}.
\item \texttt{--verbose} \textit{integer}: Determines the level of verbosity of the program (default 1).
\item \texttt{--help}: Prints information about all available options of \zebra{}.
\end{description}
Many of the aforementioned options have a single character abbreviation. Please call \texttt{zebra -h} to see the complete list of options available. 

\section{The different run modi of \lzebra{}}
\label{sect:runModes}

\zebra{} can be run in various operation modi which are specified with the run-mode options in table \ref{tab:opModi}. 

In principle it is possible to start \zebra{} with multiple run-mode options activated (except for {\it template-improvement mode}). In this case \zebra{} executes the different modi sequentially in the following order {\it k-correction mode}, {\it photometry-check mode}, {\it Maximum-Likelihood mode}, {\it Bayesian mode}. The use of multiple run-mode options is not recommended.

\begin{table}
\begin{tabular}{|p{40mm}|p{50mm}|p{52mm}|}
\hline
run mode & what it does & command line option(s) \\
\hline\hline

{\it photometry-check mode} & calibrates the photometry of a given photometric catalog & 
\mbox{\texttt{--calc-catalog-correction}} \mbox{\texttt{--apply-catalog-correction}} \texttt{\footnotesize--iterations-catalog-correction} \\ 
& & \\
 
{\it template-optimization mode} & optimizes the given templates using a training sample 
of galaxies with known redshifts & \mbox{\texttt{--improve-template}} \\
& & \\

{\it Maximum-Likelihood mode} & performs a Maximum-Likelihood fit for each galaxy in the catalog 
using the given templates. & \mbox{\texttt{--calcLikelihood}} (or \texttt{-k})\\
& & \\

{\it Bayesian mode} & estimates a prior from the catalog which is used to calculate the posterior for
each galaxy. & \mbox{\texttt{--calcBayesian}} (or \texttt{-b})\\
& & \\

{\it k-correction mode} & derives template-based k-corrections using the specified templates and 
filters. & \mbox{\texttt{--kcorrection}} \\
\hline
\end{tabular}
\caption{The different run modi of \zebra{}.}
\label{tab:opModi}
\end{table}

A typical usage of \zebra{} is indicated in the work-flow diagram fig.~\ref{workflow}.
The first step is to calibrate the photometry, i.e. running \zebra{} in {\it photometry-check mode}.
Subsequently the given templates may be improved. The template improvement reduces
systematic under/-overpredictions and the mean error of the redshift estimates.
It is strongly recommended to use a training sample with \emph{spectroscopic}
redshifts in this mode (option \texttt{--rightz} has to be enabled).
With a calibrated catalog and appropriate templates at hand \zebra{} can be run in 
{\it Maximum-Likelihood mode} and/or {\it Bayesian mode} on the photometric catalog to return final estimates 
for redshift, template type and other parameters.

The {\it k-correction mode} is an unrelated operation mode which allows to calculate k-corrections from the
given templates. No photometric catalog is needed in this case.

\subsection{Running the photometry-check mode}
\label{sect:PhotometryCheck}
This mode calibrates the photometry of a given catalog. Please refer to \cite{feldmann06} for more details.

\paragraph{Input:}
\begin{itemize}
\item catalog file (section \ref{sect:CatalogFile});
\item filter and template files (section \ref{sect:filterFiles}, \ref{sect:templateFiles});
\item standard configuration files (section \ref{sect:StandardConfFiles});
\item \textbf{configuration file of catalog correction}, see below
\end{itemize} 

\paragraph{Options and Explanations:}
The photometry correction consists of two steps. The first step is the regression.
\begin{enumerate}
\item For each filter band $n$ in question the flux residual $\Delta{}n$ 
between observed $f_n$ and template-based flux $f_n^T$ is calculated for each catalog entry.
\item The magnitude range of $n$ is divided into different magnitude bins $n_i$ and a constant or 
linear regression is performed on the data set $n_i$, $\Delta{}n_i$. 
\end{enumerate}
The second step is the application of the corrections to the filter bands of a catalog. 
Let $f_n$ be the flux of a catalog entry in filter band $n$ and $\Delta{}n(f_n)$ be the 
result of the regression, then $f_n\rightarrow{}f_n-\Delta{}n(f_n)$.

The following options are relevant
\begin{description}
\item \texttt{--calc-catalog-correction}: This enables the regression step and calculates and outputs the 
corrections of a given catalog.
\item \texttt{--apply-catalog-correction}: If this option is enabled previously derived corrections 
are applied to a (possibly different) catalog. The output catalog has the ending \texttt{.corr}.
\item \texttt{--iterations-catalog-correction} \textit{integer}: The calculation \textbf{and} application of the catalog calibration is done in a 
single step and iteratively \textit{integer} times. All intermediate corrections and calibrated catalogs are written out.
\item \texttt{--catalog-correction-def} \textit{filename}: Defines the name of the photometry-correction configuration file which 
contains, e.g. binning information for each filter band, see below. Default is \texttt{catalogCorrection.def}.
\item \texttt{--catalog-correction-base} \textit{filename}: Defines the basename (prefix) of the file into which catalog correction data is printed. Default is catalogCorrection.
\item \texttt{--single-fit}: If enabled, each catalog entry is fitted by all provided filter bands and the flux residual in each filter band is then calculated as the difference between measured flux and best fit flux. If this option is disabled 
the flux residual in the filter band $n$ is obtained from a $\chi^2$ fit in which 
the error in band $n$ is multiplied by a certain factor (given by \mbox{\texttt{--downgrading-factor}}), thereby reducing the influence of the band $n$ to the fit. In principle this leads to a faster convergence of 
the photometry correction scheme. A downgrading factor of unity reduces this case to the correction mode with \texttt{--single-fit} enabled.
\item \texttt{--downgrading-factor} \textit{float}: The role of this factor is explained in the \texttt{--single-fit} option.
\end{description}

The configuration file \texttt{catalogCorrection.def} determines 
\begin{enumerate}
\item which filter band will be corrected;
\item how many bins in magnitude space are used and what are their limits; and
\item which regression type is applied (constant offset or linear correction).
\end{enumerate}
The structure of the file is shown in Table~\ref{tab:catalogCorrection.def}. The first column determines the full 
filter name subject to photometry correction. The next column contains the number of magnitude bins. 
This number should not be too large because otherwise some bins might not contain enough catalog entries.
The third and fourth columns specify the lower (bright) and upper (faint) magnitude limits which are used to exclude very bright or faint galaxies from the regression. The 5th column denotes the applied regression mode.
We recommend using a regression with a constant offset as first choice. The extrapolation mode determines 
how \zebra{} corrects objects with magnitudes out of the magnitude range specified in columns 3 and 4. 
The range of extrapolation is specified in the next two columns. Objects with magnitudes out of this range
are subject to the same correction as objects with the nearest magnitude still within the extrapolation range. In this 
way discontinuities at the extrapolation limits are avoided. You should  make sure that the extrapolation range is large enough if 
you employ non-constant corrections. The regression curve is sampled with the resolution given in column 9.
In column 10 the user may specify that only a certain fraction $l_Y$ of objects in 
each magnitude bin is used. The fraction $1-l_Y$ of objects with the largest standard deviation from the mean flux-residual in that bin is rejected. Column 11 sets the minimum number of entries per magnitude bin. Bins with fewer objects are not
considered in the regression. The last three columns are optional.
 
In the \textit{photometry-check mode} either spectroscopic or photometric redshifts can be employed. The use of the former 
is recommended if they are available for a sufficient large subsample of the given photometric catalog. In this case the 
option \texttt{--rightz} has to be enabled. In its iterative mode \zebra{} calculates the necessary corrections and applies them to the 
given catalog. The \texttt{--apply-catalog-correction} option provides a convenient way of applying the calculated corrections to any other 
catalog.

\begin{table}
\begin{tabular}{|c|c|c|p{50mm}|}
\hline
column & content type & content & remark\\
\hline\hline
1      & string  & filter name & \\
2      & integer  & number of magnitude bins & \\
3      & float & lower magnitude bound & the actual bins are calculated from columns 2,3 and 4\\
4      & float & upper magnitude bound & \\
5      & integer & interpolation mode & 
\mbox{constant regression (0)}
\mbox{linear regression (1)}
\mbox{regression of order \textit{n} (\textit{n})} \\
6      & integer & extrapolation mode & analogous to interpolation mode \\
7      & float & lower limit of extrapolation & optional, default: \\ 
       &       &                              & smallest magnitude found \\
8      & float & upper limit of extrapolation & optional, default:\\ 
       &       &                              & largest magnitude found \\
9      & float & resolution of magnitude binning & optional, default: 0.05 \\
10      & float & percentile rejection limit & 
\mbox{optional, default: 0.954\phantom{xxxxxxxxxxxxxx}} 
\mbox{reject all (0)\phantom{xxxxxxxxxxxxxx}}
\mbox{reject none (1)\phantom{xxxxxxxxxxxxxx}} 
\mbox{use median, not mean ($<0$)\phantom{xxxxxxxxxxxxxx}}\\
11    & integer & min. number of galaxies per mag. bin & optional, default: 10 \\ 
\hline
\end{tabular}
\caption{The structure of the catalog-correction configuration file where columns 1-5 are the most important columns.
Each line corresponds to a different filter band. Comment lines have to start with \#. Only for filter bands
which are mentioned here a photometry correction is performed. As interpolation mode we recommend constant (0) or linear regression (1).}
\label{tab:catalogCorrection.def}
\end{table}

\paragraph{Output:}
\zebra{} creates the following files in the \textit{photometry-check mode}:
\begin{itemize}
\item \texttt{ML.dat}: This file summarizes the Maximum-Likelihood fit results \textbf{before} the photometry correction. If an iterative photometry correction is
 performed \texttt{ML\_1.dat} corresponds to the Maximum-Likelihood fit on the original catalog, while \texttt{ML.dat} corresponds to the Maximum-Likelihood 
 fit run on the last-before-final corrected catalog. Hence, to obtain the Maximum-Likelihood results on the photometry-corrected 
 catalog you need to run \zebra{} in {\it Maximum-Likelihood mode} on the final photometry-corrected catalog.
\item \texttt{residual.dat}: The observed magnitude and the magnitude constructed from the best fit template is stored in this file.
\item \texttt{catalogCorrection}\_\textit{filtername}\texttt{.corr}: Stores the regression curve (and the estimated errors)
within the extrapolation limits, i.e. it contains the magnitude corrections as a function of magnitude.
\item \texttt{catalogCorrection}\_\textit{filtername}\texttt{.stat}: Contains the residuals and their errors between
observed and template-based magnitude for each magnitude bin. It also stores the regression results.
\item \textit{catalog}\texttt{.corr}: This is the calibrated catalog.
\end{itemize}
Files from intermediate iterations (not the final iteration) are distinguished by an additional \_1 suffix to the catalog correction base for the first iteration, \_2 for the second and so on.

The corrected catalogs (from \textit{catalog}\_1\texttt{.corr} until finally \textit{catalog}\texttt{.corr}) store the corrections in a cumulative way. This is \textbf{not} true
for the correction (\texttt{catalogCorrection}\_*) files.  Hence, the first iteration (i.e. the correction files with \_1) usually gives 
the largest contribution to the total photometry correction.

The format and content of the catalogs that \zebra{} outputs depend on the options \\
\texttt{--preserve-errors-output} and \texttt{--preserve-catalog-format}. When \zebra{} is run in the {\it photometry-correction mode} we recommend enabling these options.

\paragraph{Remarks:}
\begin{itemize}
\item You should make sure that there is no systematic problem with your templates. You can combine the information in \texttt{ML.dat} and \texttt{residual.dat} 
to check that your corrections hold independently of template type. If you have problems with your templates you might want to skip the photometry
correction step and proceed with the template improvement. However, without sensible photometry it is not clear whether template improvement will give you a
trustworthy result. Alternatively you can still run the photometry correction mode and apply those corrections in all further steps while keeping in mind that
in this case the corrections are internal to \zebra{} and do not have any physical meaning.
\item We find empirically that trying to correct for more filter bands than actually necessary deteriorates the quality of the estimated photometric corrections. Hence, to obtain optimal catalog corrections one should try to correct as few bands as possible.
\item Constant offsets should be the first choice for photometric corrections.
\item If there are only very few spectroscopic redshifts available it might be worthwhile 
to run the photometry correction mode on a large catalog using the internally derived photometric redshifts instead (disable
\texttt{--rightz} option).
\end{itemize}

\subsection{Running the template-optimization mode}

This mode creates optimized templates which increase the quality of redshift and template type estimates.

\paragraph{Input:}
\begin{itemize}
\item catalog file (section \ref{sect:CatalogFile});
\item filter and template files (section \ref{sect:filterFiles}, \ref{sect:templateFiles});
\item standard configuration files (section \ref{sect:StandardConfFiles});
\item \textbf{configuration file for template optimization}, as described below;
\item possibly a file specifying $\rho^2$ and $\sigma^2$ values for individual wavelength (see below);
\item possibly such a file for each template (see below)
\end{itemize} 

\paragraph{Options and Explanations:}
The following options are relevant:
\begin{description}
\item \texttt{--improve-template}: Activates the \textit{template-optimization mode}. This mode cannot be combined with
other run modi of \zebra{}.
\item \texttt{--template-improvement-def} \textit{filename}: Determines the file containing the redshift 
bins for the template correction. Default is \texttt{templateImprovement.def}
\item \texttt{--correction-steps}: Number of iterations of the template improvement scheme. 
\item \texttt{--template-improvement-mode}: Uses the original templates as starting point of each iteration 
(0 - default) or takes the corrected templates (1). In the latter case the result depends on the iteration number and
finally converges to the result for $\sigma\rightarrow{}\infty$.
\item \texttt{--sigma} \textit{float}: The smaller $\sigma$ the higher the penalty for large deviations of the corrected templates from the old templates.
\item \texttt{--rho} \textit{float}: The smaller $\rho$ the stronger is the regularization of the shape of the corrected templates (i.e. large gradients are increasingly suppressed as $\rho$ becomes smaller).
\item \texttt{--load-rho-sigma}: Indicates to use different $\rho$ and $\sigma$ values depending on wavelength and template type. The user can provide a global file with
\texttt{--rho-sigma-load-base} or individual files for each template (see the description of the template configuration 
file in section \ref{sect:StandardConfFiles}). This option is not activated by default.
\item \texttt{--save-rho-sigma}: Save a file containing $\rho^2$ and $\sigma^2$ values depending on wavelength. The output filename can be specified 
with the \texttt{--rho-sigma-save-base} option. This option is not activated by default.
\item \texttt{--punish-gradient}: If this option is not activated, $\rho$ is set to infinity, independently of the choice of the option \texttt{--rho}.
\item \texttt{--rescale-with-mesh}: $\sigma$ and $\rho$ are rescaled with the square root and inverse square root of the grid spacing to remove dependencies on the mesh-spacing.
\item \texttt{--rescale-with-flux} \textit{integer}: $\sigma$ and $\rho$ are rescaled with the template flux to avoid over-corrections in the low-flux regions. Three different modi are available. (0) no rescaling, (1)  $\sigma$ and $\rho$ scale with the square root of the flux, (2)  $\sigma$ and $\rho$ scale proportionally with the flux. We recommend option (1).
\item \texttt{--rescale-with-bands}: This option can be used to reduce the influence of the number of filter bands on the outcome of the template improvement. However, depending on this option, different $\rho$ and $\sigma$ values are required.
\end{description}

The configuration file \texttt{templateImprovement.def} defines the limits of the redshift bins for the correction. Each redshift bin is
handled by \zebra{} in an independent manner.  The file contains only one column and needs to consist of at least two rows. The row values have to increase monotonically from top row to bottom row. The first row determines the low-redshift limit of the first redshift bin, the second row specifies the high-redshift limit of the first redshift bin and simultaneously the low-redshift limit of the second bin etc. 

In addition to global $\rho$ and $\sigma$ values which hold for all templates and wavelengths the user may specify (using the option \texttt{--load-rho-sigma})
\begin{itemize}
\item a file containing $\sigma^2$ and $\rho^2$ values for individual wavelengths (default \texttt{rhoSigma.dat}),
\item such a file specific for each template.
\end{itemize}
Each of those files needs to consist of at least two columns containing $\sigma^2$ and $\rho^2$, respectively. The number of rows has to be identical to the size of the wavelength mesh used in the template optimization. The easiest way to create such a file is to first to run \zebra{} with the option 
\texttt{--save-rho-sigma} (and \texttt{--load-rho-sigma} disabled) and then to copy/modify the resulting output file.

\paragraph{Output:}
The following files are written when \zebra{} is run in {\it template-optimization mode}:
\begin{itemize}
\item \texttt{Template}\_\texttt{corr}\_\texttt{it}\textit{I}\_l\textit{T}\texttt{.dat}: Contains the spectral energy distribution of template number
\textit{T} after \textit{I}$+1$ iteration steps. All templates, i.e. the original templates, the corrected templates and all interpolations
are included in this enumeration. 
\item \textit{template-filename}\_\texttt{corr}\_\texttt{it}\textit{I}\_\texttt{zgt}\textit{zlow}\_\texttt{zlt}\textit{zhigh}\texttt{.dat}: Contains the spectral energy 
distribution of the corrected basic template with original name \textit{template-filename} after \textit{I}$+1$ 
iterations which has been optimized for the redshift range \textit{zlow}--\textit{zhigh}.
\end{itemize}

The option \texttt{--rightz} is \textbf{not} automatically activated when this operation mode is used.

\paragraph{Remarks:}
\begin{itemize}
\item The optimal choice of $\sigma$ and $\rho$ depends on the data at hand. In any case you should look at the improved templates and judge whether their
shapes are sensible or not. Extreme wiggles may indicate that $\rho$ is too high, while too strong changes of the templates can originate from a
too high $\sigma$. In addition, having too few catalog entries (for a given redshift bin and given type) can imply non-physical template shapes due to
over-fitting.
\item You should check that your redshift binning scheme does not lead to artificial peaks and voids
in the redshift distribution near the bin boundaries. In order to minimize such spurious effects we
recommend running the {\it template-optimization mode} twice, with interlaced redshift binning
schemes, and then to combine the optimized templates.
\end{itemize}

\subsection{Running the  Maximum-Likelihood mode}
\label{sect:RunML}

This mode estimates redshift, template type and other parameters using a Maximum-Likelihood approach.

\paragraph{Input:}
\begin{itemize}
\item catalog file (section \ref{sect:CatalogFile});
\item filter and template files (section \ref{sect:filterFiles}, \ref{sect:templateFiles});
\item standard configuration files (section \ref{sect:StandardConfFiles});
\end{itemize} 

\paragraph{Options and Explanations:}
The following options are relevant:
\begin{description}
\item \texttt{--calcLikelihood} or \texttt{-k}: Runs \zebra{} in {\it Maximum-Likelihood mode}.
\item \texttt{--likelihood full}: Saves the 2D (redshift and template type) likelihood function $L(z,t)$ for
each galaxy
\item \texttt{--likelihood t}: Saves the marginalized likelihood function $L(t)$.
\item \texttt{--likelihood z}: Saves the marginalized likelihood function $L(z)$.
\item \texttt{--likelihood pz}: Saves the marginalized likelihood function $L(z)$ in form of percentiles.
\item \texttt{--dump besttemplate}: Saves for each galaxy the best fit template. The template is
normalized to best fit the observational data.
\item \texttt{--dump templates}: Saves for each galaxy all fitted templates (for diagnostics). The templates are
normalized to best fit the observational data.
\item \texttt{--dump fluxes}: Saves the fluxes in each filter band of each galaxy.
\item \texttt{--dump residual}: Saves the observed and template-based flux in magnitudes in each filter band
for each galaxy.
\item \texttt{--single}: If activated all the output specified with \texttt{--likelihood} and 
\texttt{--dump} is printed into a single file. Otherwise individual files are generated for each galaxy.
\end{description}

\paragraph{Ouput:}
The results of the \textit{Maximum-Likelihood mode} are stored in the file \texttt{ML.dat}. This file is described in 
detail in section \ref{sect:OutputFiles}. Additional output files can be specified with the 
\texttt{--likelihood} or \texttt{--dump} option. Their layout is explained by a header within each file.

\subsection{Running the Bayesian mode}

This mode estimates redshift, template type and other parameters using a Bayesian approach. The prior
is calculated in a self-consistent way from the given photometric catalog.

\paragraph{Input:}
\begin{itemize}
\item catalog file (section \ref{sect:CatalogFile});
\item filter and template files (section \ref{sect:filterFiles}, \ref{sect:templateFiles});
\item standard configuration files (section \ref{sect:StandardConfFiles});
\end{itemize} 

\paragraph{Options and Explanations:}
The following options are relevant:
\begin{description}
\item \texttt{--calcBayesian} or \texttt{-k}: Runs \zebra{} in {\it Bayesian mode}.
\item \texttt{--calc-prior}: The prior is calculated iteratively (starting with a flat or loaded prior).
\item \texttt{--calc-prior-mode} \textit{integer}: Calculate prior in redshift and template space (0) or taking projected
prior using the best fit template (1).
\item \texttt{--max-iterations} \textit{integer}: The number of prior calculation steps.
\item \texttt{--smoothprior}: Prior is smoothed after each iteration. The size of the smoothing kernel is set with
\texttt{--smootht} and \texttt{--smoothz}.
\item \texttt{--smootht} \textit{float}: FWHM of rectangular smoothing kernel per template in template space. \textit{float} has to be non-positive if no smoothing is wanted.
\item \texttt{--smoothz} \textit{float}: FWHM of rectangular smoothing kernel in red shift space. \textit{float} has to be non-positive if no smoothing is wanted.s
\item \texttt{--smoothz-mode} \textit{integer}: Smoothing mode for prior in redshift space. Possible choices are a fixed
kernel width (0) or an adaptive width (1), i.e. scaling with $(1+z)$.
\item \texttt{--load-prior}: Prior is read from file (specified with \texttt{--prior-load-base}).
\item \texttt{--save-prior}: Calculated prior saved into file (specified with \texttt{--priorbase}).
\item \texttt{--save-prior-iterative}: Prior is saved after each iterative step.
\item \texttt{--priorbase} \textit{filename}: Basename (prefix) of file storing the prior $P(z,t)$.
\item \texttt{--prior-load-base} \textit{filename}: Basename (prefix) of file containing the prior to load.
\item \texttt{--priorpath} \textit{pathname}: Directory containing the prior file.
\item \texttt{--posterior full}: Saves the 2D-posterior $P(z,t)$ for each galaxy.
\item \texttt{--posterior t}: Saves the posterior $P(t)$ marginalized over redshifts.
\item \texttt{--posterior z}: Saves the posterior $P(z)$ marginalized over template types.
\item \texttt{--posterior pz}: Saves the posterior $P(z)$ marginalized over template types as percentiles.
 \item \texttt{--single}: If activated, all the output specified with \texttt{--posterior} is printed 
into a single file. Otherwise individual files are generated for each galaxy.
\end{description}

\paragraph{Ouput:}
The results of the \textit{Bayesian mode} are stored in the file \texttt{Bayes\_ML.dat}. This file is described in detail in \ref{sect:OutputFiles}.
Additional output files can be specified with the \texttt{--posterior} option. Their layout is explained by a header 
within each file.

\subsection{Running the k-correction mode}

In this run mode \zebra{} returns k-corrections based on the given templates and filter bands.

\paragraph{Input:}
\begin{itemize}
\item filter and template files (section \ref{sect:filterFiles}, \ref{sect:templateFiles});
\item filter and template configuration files (section \ref{sect:StandardConfFiles});
\end{itemize} 

\paragraph{Options:}
The \textit{k-correction mode} is chosen by the option \texttt{--kcorrection}.

\paragraph{Output:}
The following output files are created.
\begin{itemize}
\item kCorrection.dat: Contains magnitudes for each template in each filter band as a function of redshift. The flux is not normalized as only flux ratios, i.e. magnitude differences, are of interest.
\item kCorrection\_real.dat: Contains for each template the k-Correction (as a function of redshift) according to the
definition of \cite{Hogg:2002yh}. The used formula is also stated in the header of this file. The first filter in the filter
configuration file defines the rest-frame band, the second filter the observed-frame filter band.
\item kCorrection\_zero.dat: Contains the so-called color-k-correction for each template as a function of redshift. The applied formula is stated in the header of this file. The first filter in the filter
configuration file defines the rest-frame filter band, the second filter the observed-frame filter band.
\end{itemize}

\section{Background Information}

\subsection{Extensions of templates and the mesh}
\zebra{} complains if a template is too short on either the short- or long-wavelength side as described in \ref{sect:templateFiles}. The user has then either to enlarge the template appropriately or to reduce the redshift range in question. An additional complication arises if \zebra{} is run in {\it template-improvement mode} or if the user 
enables the \texttt{--use-mesh} option. In this case each template and filter is re-sampled on a mesh. If the mesh contains less wavelength sampling points than the original templates \zebra{} will run faster. The user should ensure, however, to use a high enough resolution (this depends also in the filters used). 
The properties of the mesh can be set with the options \texttt{mesh-lambda-min}, \texttt{mesh-lambda-max} and \texttt{mesh-lambda-res}. The mesh has to be included in the original template range, otherwise \zebra{} will output a warning. In conclusion, the wavelength range of the mesh has to be included in the wavelength range of all templates, but it has to be large enough that fluxes in all filter bands at all considered redshifts can be calculated.

\subsection{The use of cosmological parameters}
The cosmological parameters $\Omega_M$, $\Omega_\Lambda$ and $h$ in 100 km/s/Mpc are (only) used to transform apparent
magnitudes into absolute magnitudes, e.g. to ensure the absolute magnitude cut. The user may change the default values
with the \texttt{--omegaM}, \texttt{--omegaL} and \texttt{--hubble} options.
Calling \zebra{} with higher verbosity
(\texttt{-v 3}) will print (among other information) a table containing the luminosity distance $D_L$ on the screen.
The cosmology does not need to be a flat one, i.e. \zebra{} assumes $\Omega_k=1 - \Omega_M - \Omega_\Lambda$.

\subsection{The intergalactic absorption}
\label{sect:IGA}
\begin{figure}
\includegraphics[bb=25 130 550 635, width=160mm]{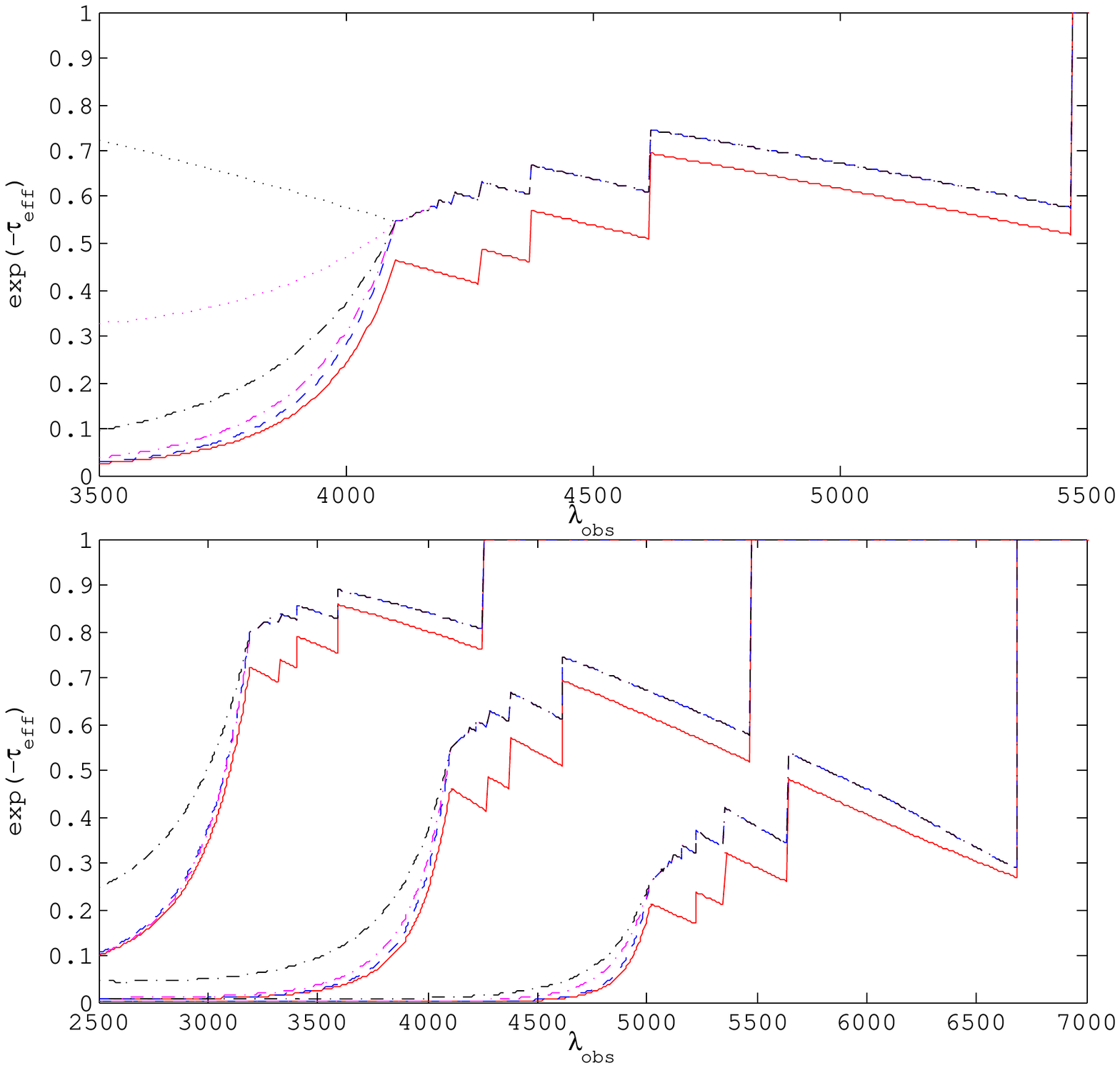}
\caption{Mean intergalactic transmission vs observed wavelength. (Top) For an object at 
redshift 3.5 and for different attenuation recipes enumerated by the value of the 
\texttt{ig-absorption} parameter (see text):
1 (red solid line), 2 (blue dashed line), 3 (magenta dot-dashed line), 4 (black dot-dashed line),
5 (magenta dotted line), 6 (black dotted line). (Bottom) As in the top panel, but for objects at redshift 
2.5 (left), 3.5 (middle), 4.5 (right).}
\label{fig:IGA}
\end{figure}

\zebra{} can take the attenuation of templates at high redshifts due to intervening
intergalactic hydrogen into account. The attenuation is mainly be caused by resonant scattering of 
Lyman-series transitions (LT) and by photoelectic absorption due to optically thin (OT) or Lyman-limit (LL) 
systems. The parameter \texttt{ig-absorption} controls the employed attenuation recipe: (0) no
attenuation is used, (1) attenuation (LT+OT+LL) according to \cite{1995ApJ...441...18M} 
(considering L$_\alpha$ - L$_\delta$ transitions), (2) LT absorption according 
to \cite{2006MNRAS.365..807M} (considering the first 40 line transitions) and OT+LL according 
to \cite{1995ApJ...441...18M}, (3) attenuation (LT+OT+LL) according to \cite{2006MNRAS.365..807M}, (4) attenuation (LT+OT+LL) according to
\cite{MeiksinPrivComm}, (5) attenuation (LT+OT) according to \cite{MeiksinPrivComm}, (6)
only LT attenuation according to \cite{2006MNRAS.365..807M}. The different attenuation recipes
implemented in \zebra{} are shown in Fig~\ref{fig:IGA} and should be compared with Fig~3 of \cite{1995ApJ...441...18M} and Fig~1
of \cite{2006MNRAS.365..807M}.

\subsection{Flux densities and normalization}
\zebra{} operates internally on flux densities (per unit frequency). Catalog entries are given in 
magnitudes and are converted to flux densities as described in \ref{sect:CatalogFile}.
The flux density (per unit frequency) in a certain filter band of a red-shifted template is calculated 
according to eq. A3 of \cite{feldmann06}. This is done \emph{before} any Maximum-Likelihood fit is performed and thus independent of the 
actual values of the provided catalog. The flux-densities can be saved and/or loaded with 
the \texttt{--save-flux} and \texttt{--load-flux} options in order to save time in case \zebra{} is run multiple 
times with the same set of templates, filters and redshift binning. 

You can use the saved fluxes to calculate
the template-based magnitudes for each galaxy in each filter (although there are also other ways to do this, see
\ref{sect:RunML}, in particular options \texttt{--dump residual} and \texttt{--dump fluxes}). You need to extract 
the flux corresponding to the best-fit template and the best-fit redshift of the desired filter 
band from the saved flux-file and multiply it with the best-fit normalization. Then you convert this flux density into 
a magnitude according to \ref{sect:CatalogFile}. Best-fit redshift (column 3), best-fit 
template (column 4) and best-fit normalization (column 5) can be found in \texttt{ML.dat}.
If you want to convert a flux density per unit frequency $f_\nu$ into a flux density per unit wavelength
$f_\lambda$ you need to divide $f_\nu$ by the square of the pivot-wavelength of each filter. The
pivot-wavelengths of the filters are printed (among other information) if 
\zebra{} is run with \texttt{--verbose 2}.

\zebra{} offers the possibility of normalizing input templates to the dedicated B-filter using the \texttt{--normalize-templates} option. 
Its application can change the values in the flux file and 
the best-fit normalization but not the best-fit template or the best-fit redshift unless linearly interpolated
templates are used. If only logarithmically interpolated templates are employed the normalization of the
template files has no influence on the result of a Maximum-Likelihood or Bayesian run.

\section{Experimental features}

Several features are implemented in the current version of \zebra{} which are not explained in this manual.
They are neither necessarily well tested nor fully functional and you should avoid using them. They are:
\begin{itemize}
\item \texttt{--chi2-add}
\item \texttt{--chi2-mult}
\item \texttt{--lin-interpolation}
\item \texttt{--improve-template-mode 1}
\item \texttt{--smoothz-mode 2}
\end{itemize}

\vspace{0.2cm}

\bibliography{docZEBRA}

\begin{thebibliography}{7}
\expandafter\ifx\csname natexlab\endcsname\relax\def\natexlab#1{#1}\fi

\bibitem[{{Coleman} {et~al.}(1980){Coleman}, {Wu}, \& {Weedman}}]{cww}
{Coleman}, G.~D., {Wu}, C.-C., \& {Weedman}, D.~W. 1980, \apjs, 43, 393

\bibitem[{{Feldmann} {et~al.}(2006){Feldmann}, {Carollo}, {Porciani}, {Lilly},
  {Capak}, {Taniguchi}, {F{\`e}vre}, {Renzini}, {Scoville}, {Ajiki}, {Aussel},
  {Contini}, {McCracken}, {Mobasher}, {Murayama}, {Sanders}, {Sasaki},
  {Scarlata}, {Scodeggio}, {Shioya}, {Silverman}, {Takahashi}, {Thompson}, \&
  {Zamorani}}]{feldmann06}
{Feldmann}, R., {Carollo}, C.~M., {Porciani}, C., {Lilly}, S.~J., {Capak}, P.,
  {Taniguchi}, Y., {F{\`e}vre}, O.~L., {Renzini}, A., {Scoville}, N., {Ajiki},
  M., {Aussel}, H., {Contini}, T., {McCracken}, H., {Mobasher}, B., {Murayama},
  T., {Sanders}, D., {Sasaki}, S., {Scarlata}, C., {Scodeggio}, M., {Shioya},
  Y., {Silverman}, J., {Takahashi}, M., {Thompson}, D., \& {Zamorani}, G. 2006,
  \mnras, 372, 565

\bibitem[{{Hogg} {et~al.}(2002){Hogg}, {Baldry}, {Blanton}, \&
  {Eisenstein}}]{Hogg:2002yh}
{Hogg}, D.~W., {Baldry}, I.~K., {Blanton}, M.~R., \& {Eisenstein}, D.~J. 2002,
  {The K correction}, arXiv:astro-ph/0210394

\bibitem[{{Kinney} {et~al.}(1996){Kinney}, {Calzetti}, {Bohlin}, {McQuade},
  {Storchi-Bergmann}, \& {Schmitt}}]{kin}
{Kinney}, A.~L., {Calzetti}, D., {Bohlin}, R.~C., {McQuade}, K.,
  {Storchi-Bergmann}, T., \& {Schmitt}, H.~R. 1996, \apj, 467, 38

\bibitem[{{Madau}(1995)}]{1995ApJ...441...18M}
{Madau}, P. 1995, \apj, 441, 18

\bibitem[{{Meiksin}(2006)}]{2006MNRAS.365..807M}
{Meiksin}, A. 2006, \mnras, 365, 807

\bibitem[{Meiksin(2007)}]{MeiksinPrivComm}
Meiksin, A. 2007, private communications

\end{thebibliography}

\clearpage

\appendix

\clearpage
\section{GNU GENERAL PUBLIC LICENSE}
GNU GENERAL PUBLIC LICENSE

\tiny
\parindent=0.1in
Version 3, 29 June 2007

Copyright (C) 2007 Free Software Foundation, Inc. \verb|<http://fsf.org/>|

 Everyone is permitted to copy and distribute verbatim copies
 of this license document, but changing it is not allowed.

                            Preamble

  The GNU General Public License is a free, copyleft license for
software and other kinds of works.

  The licenses for most software and other practical works are designed
to take away your freedom to share and change the works.  By contrast,
the GNU General Public License is intended to guarantee your freedom to
share and change all versions of a program--to make sure it remains free
software for all its users.  We, the Free Software Foundation, use the
GNU General Public License for most of our software; it applies also to
any other work released this way by its authors.  You can apply it to
your programs, too.

  When we speak of free software, we are referring to freedom, not
price.  Our General Public Licenses are designed to make sure that you
have the freedom to distribute copies of free software (and charge for
them if you wish), that you receive source code or can get it if you
want it, that you can change the software or use pieces of it in new
free programs, and that you know you can do these things.

  To protect your rights, we need to prevent others from denying you
these rights or asking you to surrender the rights.  Therefore, you have
certain responsibilities if you distribute copies of the software, or if
you modify it: responsibilities to respect the freedom of others.

  For example, if you distribute copies of such a program, whether
gratis or for a fee, you must pass on to the recipients the same
freedoms that you received.  You must make sure that they, too, receive
or can get the source code.  And you must show them these terms so they
know their rights.

  Developers that use the GNU GPL protect your rights with two steps:
(1) assert copyright on the software, and (2) offer you this License
giving you legal permission to copy, distribute and/or modify it.

  For the developers' and authors' protection, the GPL clearly explains
that there is no warranty for this free software.  For both users' and
authors' sake, the GPL requires that modified versions be marked as
changed, so that their problems will not be attributed erroneously to
authors of previous versions.

  Some devices are designed to deny users access to install or run
modified versions of the software inside them, although the manufacturer
can do so.  This is fundamentally incompatible with the aim of
protecting users' freedom to change the software.  The systematic
pattern of such abuse occurs in the area of products for individuals to
use, which is precisely where it is most unacceptable.  Therefore, we
have designed this version of the GPL to prohibit the practice for those
products.  If such problems arise substantially in other domains, we
stand ready to extend this provision to those domains in future versions
of the GPL, as needed to protect the freedom of users.

  Finally, every program is threatened constantly by software patents.
States should not allow patents to restrict development and use of
software on general-purpose computers, but in those that do, we wish to
avoid the special danger that patents applied to a free program could
make it effectively proprietary.  To prevent this, the GPL assures that
patents cannot be used to render the program non-free.

  The precise terms and conditions for copying, distribution and
modification follow.

                       TERMS AND CONDITIONS

  0. Definitions.

  "This License" refers to version 3 of the GNU General Public License.

  "Copyright" also means copyright-like laws that apply to other kinds of
works, such as semiconductor masks.

  "The Program" refers to any copyrightable work licensed under this
License.  Each licensee is addressed as "you".  "Licensees" and
"recipients" may be individuals or organizations.

  To "modify" a work means to copy from or adapt all or part of the work
in a fashion requiring copyright permission, other than the making of an
exact copy.  The resulting work is called a "modified version" of the
earlier work or a work "based on" the earlier work.

  A "covered work" means either the unmodified Program or a work based
on the Program.

  To "propagate" a work means to do anything with it that, without
permission, would make you directly or secondarily liable for
infringement under applicable copyright law, except executing it on a
computer or modifying a private copy.  Propagation includes copying,
distribution (with or without modification), making available to the
public, and in some countries other activities as well.

  To "convey" a work means any kind of propagation that enables other
parties to make or receive copies.  Mere interaction with a user through
a computer network, with no transfer of a copy, is not conveying.

  An interactive user interface displays "Appropriate Legal Notices"
to the extent that it includes a convenient and prominently visible
feature that (1) displays an appropriate copyright notice, and (2)
tells the user that there is no warranty for the work (except to the
extent that warranties are provided), that licensees may convey the
work under this License, and how to view a copy of this License.  If
the interface presents a list of user commands or options, such as a
menu, a prominent item in the list meets this criterion.

  1. Source Code.

  The "source code" for a work means the preferred form of the work
for making modifications to it.  "Object code" means any non-source
form of a work.

  A "Standard Interface" means an interface that either is an official
standard defined by a recognized standards body, or, in the case of
interfaces specified for a particular programming language, one that
is widely used among developers working in that language.

  The "System Libraries" of an executable work include anything, other
than the work as a whole, that (a) is included in the normal form of
packaging a Major Component, but which is not part of that Major
Component, and (b) serves only to enable use of the work with that
Major Component, or to implement a Standard Interface for which an
implementation is available to the public in source code form.  A
"Major Component", in this context, means a major essential component
(kernel, window system, and so on) of the specific operating system
(if any) on which the executable work runs, or a compiler used to
produce the work, or an object code interpreter used to run it.

  The "Corresponding Source" for a work in object code form means all
the source code needed to generate, install, and (for an executable
work) run the object code and to modify the work, including scripts to
control those activities.  However, it does not include the work's
System Libraries, or general-purpose tools or generally available free
programs which are used unmodified in performing those activities but
which are not part of the work.  For example, Corresponding Source
includes interface definition files associated with source files for
the work, and the source code for shared libraries and dynamically
linked subprograms that the work is specifically designed to require,
such as by intimate data communication or control flow between those
subprograms and other parts of the work.

  The Corresponding Source need not include anything that users
can regenerate automatically from other parts of the Corresponding
Source.

  The Corresponding Source for a work in source code form is that
same work.

  2. Basic Permissions.

  All rights granted under this License are granted for the term of
copyright on the Program, and are irrevocable provided the stated
conditions are met.  This License explicitly affirms your unlimited
permission to run the unmodified Program.  The output from running a
covered work is covered by this License only if the output, given its
content, constitutes a covered work.  This License acknowledges your
rights of fair use or other equivalent, as provided by copyright law.

  You may make, run and propagate covered works that you do not
convey, without conditions so long as your license otherwise remains
in force.  You may convey covered works to others for the sole purpose
of having them make modifications exclusively for you, or provide you
with facilities for running those works, provided that you comply with
the terms of this License in conveying all material for which you do
not control copyright.  Those thus making or running the covered works
for you must do so exclusively on your behalf, under your direction
and control, on terms that prohibit them from making any copies of
your copyrighted material outside their relationship with you.

  Conveying under any other circumstances is permitted solely under
the conditions stated below.  Sublicensing is not allowed; section 10
makes it unnecessary.

  3. Protecting Users' Legal Rights From Anti-Circumvention Law.

  No covered work shall be deemed part of an effective technological
measure under any applicable law fulfilling obligations under article
11 of the WIPO copyright treaty adopted on 20 December 1996, or
similar laws prohibiting or restricting circumvention of such
measures.

  When you convey a covered work, you waive any legal power to forbid
circumvention of technological measures to the extent such circumvention
is effected by exercising rights under this License with respect to
the covered work, and you disclaim any intention to limit operation or
modification of the work as a means of enforcing, against the work's
users, your or third parties' legal rights to forbid circumvention of
technological measures.

  4. Conveying Verbatim Copies.

  You may convey verbatim copies of the Program's source code as you
receive it, in any medium, provided that you conspicuously and
appropriately publish on each copy an appropriate copyright notice;
keep intact all notices stating that this License and any
non-permissive terms added in accord with section 7 apply to the code;
keep intact all notices of the absence of any warranty; and give all
recipients a copy of this License along with the Program.

  You may charge any price or no price for each copy that you convey,
and you may offer support or warranty protection for a fee.

  5. Conveying Modified Source Versions.

  You may convey a work based on the Program, or the modifications to
produce it from the Program, in the form of source code under the
terms of section 4, provided that you also meet all of these conditions:

    a) The work must carry prominent notices stating that you modified
    it, and giving a relevant date.

    b) The work must carry prominent notices stating that it is
    released under this License and any conditions added under section
    7.  This requirement modifies the requirement in section 4 to
    "keep intact all notices".

    c) You must license the entire work, as a whole, under this
    License to anyone who comes into possession of a copy.  This
    License will therefore apply, along with any applicable section 7
    additional terms, to the whole of the work, and all its parts,
    regardless of how they are packaged.  This License gives no
    permission to license the work in any other way, but it does not
    invalidate such permission if you have separately received it.

    d) If the work has interactive user interfaces, each must display
    Appropriate Legal Notices; however, if the Program has interactive
    interfaces that do not display Appropriate Legal Notices, your
    work need not make them do so.

  A compilation of a covered work with other separate and independent
works, which are not by their nature extensions of the covered work,
and which are not combined with it such as to form a larger program,
in or on a volume of a storage or distribution medium, is called an
"aggregate" if the compilation and its resulting copyright are not
used to limit the access or legal rights of the compilation's users
beyond what the individual works permit.  Inclusion of a covered work
in an aggregate does not cause this License to apply to the other
parts of the aggregate.

  6. Conveying Non-Source Forms.

  You may convey a covered work in object code form under the terms
of sections 4 and 5, provided that you also convey the
machine-readable Corresponding Source under the terms of this License,
in one of these ways:

    a) Convey the object code in, or embodied in, a physical product
    (including a physical distribution medium), accompanied by the
    Corresponding Source fixed on a durable physical medium
    customarily used for software interchange.

    b) Convey the object code in, or embodied in, a physical product
    (including a physical distribution medium), accompanied by a
    written offer, valid for at least three years and valid for as
    long as you offer spare parts or customer support for that product
    model, to give anyone who possesses the object code either (1) a
    copy of the Corresponding Source for all the software in the
    product that is covered by this License, on a durable physical
    medium customarily used for software interchange, for a price no
    more than your reasonable cost of physically performing this
    conveying of source, or (2) access to copy the
    Corresponding Source from a network server at no charge.

    c) Convey individual copies of the object code with a copy of the
    written offer to provide the Corresponding Source.  This
    alternative is allowed only occasionally and noncommercially, and
    only if you received the object code with such an offer, in accord
    with subsection 6b.

    d) Convey the object code by offering access from a designated
    place (gratis or for a charge), and offer equivalent access to the
    Corresponding Source in the same way through the same place at no
    further charge.  You need not require recipients to copy the
    Corresponding Source along with the object code.  If the place to
    copy the object code is a network server, the Corresponding Source
    may be on a different server (operated by you or a third party)
    that supports equivalent copying facilities, provided you maintain
    clear directions next to the object code saying where to find the
    Corresponding Source.  Regardless of what server hosts the
    Corresponding Source, you remain obligated to ensure that it is
    available for as long as needed to satisfy these requirements.

    e) Convey the object code using peer-to-peer transmission, provided
    you inform other peers where the object code and Corresponding
    Source of the work are being offered to the general public at no
    charge under subsection 6d.

  A separable portion of the object code, whose source code is excluded
from the Corresponding Source as a System Library, need not be
included in conveying the object code work.

  A "User Product" is either (1) a "consumer product", which means any
tangible personal property which is normally used for personal, family,
or household purposes, or (2) anything designed or sold for incorporation
into a dwelling.  In determining whether a product is a consumer product,
doubtful cases shall be resolved in favor of coverage.  For a particular
product received by a particular user, "normally used" refers to a
typical or common use of that class of product, regardless of the status
of the particular user or of the way in which the particular user
actually uses, or expects or is expected to use, the product.  A product
is a consumer product regardless of whether the product has substantial
commercial, industrial or non-consumer uses, unless such uses represent
the only significant mode of use of the product.

  "Installation Information" for a User Product means any methods,
procedures, authorization keys, or other information required to install
and execute modified versions of a covered work in that User Product from
a modified version of its Corresponding Source.  The information must
suffice to ensure that the continued functioning of the modified object
code is in no case prevented or interfered with solely because
modification has been made.

  If you convey an object code work under this section in, or with, or
specifically for use in, a User Product, and the conveying occurs as
part of a transaction in which the right of possession and use of the
User Product is transferred to the recipient in perpetuity or for a
fixed term (regardless of how the transaction is characterized), the
Corresponding Source conveyed under this section must be accompanied
by the Installation Information.  But this requirement does not apply
if neither you nor any third party retains the ability to install
modified object code on the User Product (for example, the work has
been installed in ROM).

  The requirement to provide Installation Information does not include a
requirement to continue to provide support service, warranty, or updates
for a work that has been modified or installed by the recipient, or for
the User Product in which it has been modified or installed.  Access to a
network may be denied when the modification itself materially and
adversely affects the operation of the network or violates the rules and
protocols for communication across the network.

  Corresponding Source conveyed, and Installation Information provided,
in accord with this section must be in a format that is publicly
documented (and with an implementation available to the public in
source code form), and must require no special password or key for
unpacking, reading or copying.

  7. Additional Terms.

  "Additional permissions" are terms that supplement the terms of this
License by making exceptions from one or more of its conditions.
Additional permissions that are applicable to the entire Program shall
be treated as though they were included in this License, to the extent
that they are valid under applicable law.  If additional permissions
apply only to part of the Program, that part may be used separately
under those permissions, but the entire Program remains governed by
this License without regard to the additional permissions.

  When you convey a copy of a covered work, you may at your option
remove any additional permissions from that copy, or from any part of
it.  (Additional permissions may be written to require their own
removal in certain cases when you modify the work.)  You may place
additional permissions on material, added by you to a covered work,
for which you have or can give appropriate copyright permission.

  Notwithstanding any other provision of this License, for material you
add to a covered work, you may (if authorized by the copyright holders of
that material) supplement the terms of this License with terms:

    a) Disclaiming warranty or limiting liability differently from the
    terms of sections 15 and 16 of this License; or

    b) Requiring preservation of specified reasonable legal notices or
    author attributions in that material or in the Appropriate Legal
    Notices displayed by works containing it; or

    c) Prohibiting misrepresentation of the origin of that material, or
    requiring that modified versions of such material be marked in
    reasonable ways as different from the original version; or

    d) Limiting the use for publicity purposes of names of licensors or
    authors of the material; or

    e) Declining to grant rights under trademark law for use of some
    trade names, trademarks, or service marks; or

    f) Requiring indemnification of licensors and authors of that
    material by anyone who conveys the material (or modified versions of
    it) with contractual assumptions of liability to the recipient, for
    any liability that these contractual assumptions directly impose on
    those licensors and authors.

  All other non-permissive additional terms are considered "further
restrictions" within the meaning of section 10.  If the Program as you
received it, or any part of it, contains a notice stating that it is
governed by this License along with a term that is a further
restriction, you may remove that term.  If a license document contains
a further restriction but permits relicensing or conveying under this
License, you may add to a covered work material governed by the terms
of that license document, provided that the further restriction does
not survive such relicensing or conveying.

  If you add terms to a covered work in accord with this section, you
must place, in the relevant source files, a statement of the
additional terms that apply to those files, or a notice indicating
where to find the applicable terms.

  Additional terms, permissive or non-permissive, may be stated in the
form of a separately written license, or stated as exceptions;
the above requirements apply either way.

  8. Termination.

  You may not propagate or modify a covered work except as expressly
provided under this License.  Any attempt otherwise to propagate or
modify it is void, and will automatically terminate your rights under
this License (including any patent licenses granted under the third
paragraph of section 11).

  However, if you cease all violation of this License, then your
license from a particular copyright holder is reinstated (a)
provisionally, unless and until the copyright holder explicitly and
finally terminates your license, and (b) permanently, if the copyright
holder fails to notify you of the violation by some reasonable means
prior to 60 days after the cessation.

  Moreover, your license from a particular copyright holder is
reinstated permanently if the copyright holder notifies you of the
violation by some reasonable means, this is the first time you have
received notice of violation of this License (for any work) from that
copyright holder, and you cure the violation prior to 30 days after
your receipt of the notice.

  Termination of your rights under this section does not terminate the
licenses of parties who have received copies or rights from you under
this License.  If your rights have been terminated and not permanently
reinstated, you do not qualify to receive new licenses for the same
material under section 10.

  9. Acceptance Not Required for Having Copies.

  You are not required to accept this License in order to receive or
run a copy of the Program.  Ancillary propagation of a covered work
occurring solely as a consequence of using peer-to-peer transmission
to receive a copy likewise does not require acceptance.  However,
nothing other than this License grants you permission to propagate or
modify any covered work.  These actions infringe copyright if you do
not accept this License.  Therefore, by modifying or propagating a
covered work, you indicate your acceptance of this License to do so.

  10. Automatic Licensing of Downstream Recipients.

  Each time you convey a covered work, the recipient automatically
receives a license from the original licensors, to run, modify and
propagate that work, subject to this License.  You are not responsible
for enforcing compliance by third parties with this License.

  An "entity transaction" is a transaction transferring control of an
organization, or substantially all assets of one, or subdividing an
organization, or merging organizations.  If propagation of a covered
work results from an entity transaction, each party to that
transaction who receives a copy of the work also receives whatever
licenses to the work the party's predecessor in interest had or could
give under the previous paragraph, plus a right to possession of the
Corresponding Source of the work from the predecessor in interest, if
the predecessor has it or can get it with reasonable efforts.

  You may not impose any further restrictions on the exercise of the
rights granted or affirmed under this License.  For example, you may
not impose a license fee, royalty, or other charge for exercise of
rights granted under this License, and you may not initiate litigation
(including a cross-claim or counterclaim in a lawsuit) alleging that
any patent claim is infringed by making, using, selling, offering for
sale, or importing the Program or any portion of it.

  11. Patents.

  A "contributor" is a copyright holder who authorizes use under this
License of the Program or a work on which the Program is based.  The
work thus licensed is called the contributor's "contributor version".

  A contributor's "essential patent claims" are all patent claims
owned or controlled by the contributor, whether already acquired or
hereafter acquired, that would be infringed by some manner, permitted
by this License, of making, using, or selling its contributor version,
but do not include claims that would be infringed only as a
consequence of further modification of the contributor version.  For
purposes of this definition, "control" includes the right to grant
patent sublicenses in a manner consistent with the requirements of
this License.

  Each contributor grants you a non-exclusive, worldwide, royalty-free
patent license under the contributor's essential patent claims, to
make, use, sell, offer for sale, import and otherwise run, modify and
propagate the contents of its contributor version.

  In the following three paragraphs, a "patent license" is any express
agreement or commitment, however denominated, not to enforce a patent
(such as an express permission to practice a patent or covenant not to
sue for patent infringement).  To "grant" such a patent license to a
party means to make such an agreement or commitment not to enforce a
patent against the party.

  If you convey a covered work, knowingly relying on a patent license,
and the Corresponding Source of the work is not available for anyone
to copy, free of charge and under the terms of this License, through a
publicly available network server or other readily accessible means,
then you must either (1) cause the Corresponding Source to be so
available, or (2) arrange to deprive yourself of the benefit of the
patent license for this particular work, or (3) arrange, in a manner
consistent with the requirements of this License, to extend the patent
license to downstream recipients.  "Knowingly relying" means you have
actual knowledge that, but for the patent license, your conveying the
covered work in a country, or your recipient's use of the covered work
in a country, would infringe one or more identifiable patents in that
country that you have reason to believe are valid.

  If, pursuant to or in connection with a single transaction or
arrangement, you convey, or propagate by procuring conveyance of, a
covered work, and grant a patent license to some of the parties
receiving the covered work authorizing them to use, propagate, modify
or convey a specific copy of the covered work, then the patent license
you grant is automatically extended to all recipients of the covered
work and works based on it.

  A patent license is "discriminatory" if it does not include within
the scope of its coverage, prohibits the exercise of, or is
conditioned on the non-exercise of one or more of the rights that are
specifically granted under this License.  You may not convey a covered
work if you are a party to an arrangement with a third party that is
in the business of distributing software, under which you make payment
to the third party based on the extent of your activity of conveying
the work, and under which the third party grants, to any of the
parties who would receive the covered work from you, a discriminatory
patent license (a) in connection with copies of the covered work
conveyed by you (or copies made from those copies), or (b) primarily
for and in connection with specific products or compilations that
contain the covered work, unless you entered into that arrangement,
or that patent license was granted, prior to 28 March 2007.

  Nothing in this License shall be construed as excluding or limiting
any implied license or other defenses to infringement that may
otherwise be available to you under applicable patent law.

  12. No Surrender of Others' Freedom.

  If conditions are imposed on you (whether by court order, agreement or
otherwise) that contradict the conditions of this License, they do not
excuse you from the conditions of this License.  If you cannot convey a
covered work so as to satisfy simultaneously your obligations under this
License and any other pertinent obligations, then as a consequence you may
not convey it at all.  For example, if you agree to terms that obligate you
to collect a royalty for further conveying from those to whom you convey
the Program, the only way you could satisfy both those terms and this
License would be to refrain entirely from conveying the Program.

  13. Use with the GNU Affero General Public License.

  Notwithstanding any other provision of this License, you have
permission to link or combine any covered work with a work licensed
under version 3 of the GNU Affero General Public License into a single
combined work, and to convey the resulting work.  The terms of this
License will continue to apply to the part which is the covered work,
but the special requirements of the GNU Affero General Public License,
section 13, concerning interaction through a network will apply to the
combination as such.

  14. Revised Versions of this License.

  The Free Software Foundation may publish revised and/or new versions of
the GNU General Public License from time to time.  Such new versions will
be similar in spirit to the present version, but may differ in detail to
address new problems or concerns.

  Each version is given a distinguishing version number.  If the
Program specifies that a certain numbered version of the GNU General
Public License "or any later version" applies to it, you have the
option of following the terms and conditions either of that numbered
version or of any later version published by the Free Software
Foundation.  If the Program does not specify a version number of the
GNU General Public License, you may choose any version ever published
by the Free Software Foundation.

  If the Program specifies that a proxy can decide which future
versions of the GNU General Public License can be used, that proxy's
public statement of acceptance of a version permanently authorizes you
to choose that version for the Program.

  Later license versions may give you additional or different
permissions.  However, no additional obligations are imposed on any
author or copyright holder as a result of your choosing to follow a
later version.

  15. Disclaimer of Warranty.

  THERE IS NO WARRANTY FOR THE PROGRAM, TO THE EXTENT PERMITTED BY
APPLICABLE LAW.  EXCEPT WHEN OTHERWISE STATED IN WRITING THE COPYRIGHT
HOLDERS AND/OR OTHER PARTIES PROVIDE THE PROGRAM "AS IS" WITHOUT WARRANTY
OF ANY KIND, EITHER EXPRESSED OR IMPLIED, INCLUDING, BUT NOT LIMITED TO,
THE IMPLIED WARRANTIES OF MERCHANTABILITY AND FITNESS FOR A PARTICULAR
PURPOSE.  THE ENTIRE RISK AS TO THE QUALITY AND PERFORMANCE OF THE PROGRAM
IS WITH YOU.  SHOULD THE PROGRAM PROVE DEFECTIVE, YOU ASSUME THE COST OF
ALL NECESSARY SERVICING, REPAIR OR CORRECTION.

  16. Limitation of Liability.

  IN NO EVENT UNLESS REQUIRED BY APPLICABLE LAW OR AGREED TO IN WRITING
WILL ANY COPYRIGHT HOLDER, OR ANY OTHER PARTY WHO MODIFIES AND/OR CONVEYS
THE PROGRAM AS PERMITTED ABOVE, BE LIABLE TO YOU FOR DAMAGES, INCLUDING ANY
GENERAL, SPECIAL, INCIDENTAL OR CONSEQUENTIAL DAMAGES ARISING OUT OF THE
USE OR INABILITY TO USE THE PROGRAM (INCLUDING BUT NOT LIMITED TO LOSS OF
DATA OR DATA BEING RENDERED INACCURATE OR LOSSES SUSTAINED BY YOU OR THIRD
PARTIES OR A FAILURE OF THE PROGRAM TO OPERATE WITH ANY OTHER PROGRAMS),
EVEN IF SUCH HOLDER OR OTHER PARTY HAS BEEN ADVISED OF THE POSSIBILITY OF
SUCH DAMAGES.

  17. Interpretation of Sections 15 and 16.

  If the disclaimer of warranty and limitation of liability provided
above cannot be given local legal effect according to their terms,
reviewing courts shall apply local law that most closely approximates
an absolute waiver of all civil liability in connection with the
Program, unless a warranty or assumption of liability accompanies a
copy of the Program in return for a fee.

                     END OF TERMS AND CONDITIONS	     
\end{document}